\documentstyle[11pt,aaspp4]{article}

\def\lya{Ly$\alpha$~}

\begin{document}

\title{Effects of Dust on Gravitational Lensing by Spiral Galaxies}

\author{Rosalba Perna and Abraham Loeb}

\affil{Harvard-Smithsonian Center for Astrophysics, MS-51\\
  60 Garden Street, Cambridge MA 02138}

\author{Matthias Bartelmann}

\affil{Max-Planck-Institut f\"ur Astrophysik\\
  Karl-Schwarzschild-Strasse 1, D--85748 Garching, Germany}

\begin{abstract}

Gravitational lensing of an optical QSO by a spiral galaxy is
often counteracted by dust obscuration, since the line-of-sight to the QSO
passes close to the center of the galactic disk. The dust in the lens is
likely to be correlated with neutral hydrogen, which in turn should leave a
\lya absorption signature on the QSO spectrum. We use the estimated
dust-to-gas ratio of the Milky--Way galaxy as a mean and allow a
spread in its values to calculate the effects of dust on lensing by
low redshift spiral galaxies.  Using a no--evolution model for spirals
at $z\la0.8$ we find, in $\Lambda=0$ cosmologies, that
the magnification bias due to lensing is stronger than dust
obscuration for QSO samples with a magnitude limit $B\la 16$.  The
density parameter of neutral hydrogen, $\Omega_{\rm HI}$, is
overestimated in such samples and is underestimated for fainter QSOs.

\end{abstract}


\centerline{ApJ, in press, 1997}

\clearpage
\section{Introduction}

Gravitational lensing by spiral galaxies results in multiple images
with sub-arcsecond separation.  A spiral galaxy at a cosmological
distance would lens a background QSO if the line-of-sight to the QSO
passed within a few kpc from the disk center.  Under these
circumstances, the neutral hydrogen in the disk would inevitably leave
a strong Ly$\alpha$ absorption signature on the quasar spectrum. A
significant HI column density was indeed inferred for the spiral
galaxy lens B0218+357 which has an Einstein ring radius $\sim 0.''3$
(Carrilli et al. 1993).

Strong \lya absorption is expected to result even at large impact
parameters from the centers of galactic disks ($\sim 10-20~{\rm kpc}$),
where the effect of lensing is weak.  In fact, damped \lya absorption
systems with neutral hydrogen column densities $\ga 10^{20}\,{\rm cm^{-2}}$
are identified routinely in the spectra of high redshift QSOs, with an
abundance consistent with that expected for the progenitors of present-day
disk galaxies (Wolfe 1995).  The association between damped \lya absorbers
and the progenitors of present-day disk galaxies is supported by several
observational facts.  First, the abundance of heavy elements at low
ionization stages in the absorbers and the velocity field traced by them is
consistent with values expected for disk galaxies (Turnshek et al.  1989;
Wolfe et al.  1993; Lu et al. 1993; Pettini et al. 1994; Lu \& Wolfe 1994).
Recent Keck HIRES observations by Wolfe (1995; see also Lanzetta \& Bowen
1992) indicate that the weak low-ionization metal absorption lines in these
systems often show the highest column-density component at either the red
or the blue edge of the velocity profile. Such an asymmetry is
characteristic of absorption by a rotating gaseous disk (Wolfe 1995).
Second, observations of redshifted 21-cm absorption and emission from
damped \lya absorbers indicate disk-like structures of galactic dimensions
(Briggs et al. 1989; Wolfe et al.  1992), and Faraday-rotation observations
are consistent with the existence of micro-Gauss magnetic fields in these
systems (Wolfe, Lanzetta, \& Oren 1992; Welter, Perry, \& Kronberg 1984;
Perry, Watson, \& Kronberg 1993).  Finally, the association of the damped
\lya absorbers with star-forming galaxies is directly confirmed by direct
imaging (Steidel et al. 1994, 1995) in most of the cases. However, whereas
some systems are clearly spirals, others show irregular morphologies (Le
Brun et al. 1996). For simplicity, we focus only on spirals. A more general
treatment of this problem would involve many more free parameters and is
beyond the scope of this paper. Neverthless, we stress that if dwarf
galaxies with relatively shallow potential wells make a significant
fraction of all damped absorbers, then our analysis overestimates the
importance of lensing.

The mass density of neutral hydrogen at $z\approx3$, $\Omega_{\rm
HI}\approx 0.005$, is comparable to the mass density in stars today,
and appears to steadily decline at lower redshifts (Lanzetta, Wolfe,
\& Turnshek 1995; Wolfe 1995 and references therein). The simplest
interpretation of this result is that the material in present-day disk
galaxies had already assembled at a redshift $z\approx 3$, while the
observed gradual decline of $\Omega_{\rm HI}(z)$ since then is simply
a result of the consumption of the gas by star formation. 

Based on these indications and the evidence that the velocity width of the
associated metal lines is indicative of rotational velocities $\sim
200~{\rm km~s^{-1}}$ (Wolfe 1995), Bartelmann \& Loeb (1996) have adopted
the view that damped Ly$\alpha$ absorption in QSO spectra results from
neutral hydrogen disks in the present-day potential wells of spiral
galaxies.  Following this assumption, they showed that damped Ly$\alpha$
systems may lens their background QSOs. An important consequence thereof is
that the statistics of the absorbers themselves, as inferred from the
number and distribution of absorption troughs detected, can be affected by
the magnification bias. Because of the magnification of their flux due to
lensing, the number of QSOs observed above a given magnitude threshold is
increased, and consequently the observed fraction of QSOs which show damped
Ly$\alpha$ absorption is enhanced.

An effect which tends to counteract the magnification by lensing is
obscuration of the source by dust. By now, there is evidence that damped
\lya systems contain measurable quantities of dust.  A dust-to-gas ratio of
order 10\% of the Milky-Way value is inferred from the enhanced level of
reddening found in a sample of QSOs that have damped \lya absorption (Fall,
Pei, \& McMahon 1989; Pei, Fall, \& Bechtold 1991), and from the gas-phase
depletion of Cr relative to Zn (see Meyer, Welty, \& York 1989; Meyer \&
Roth 1990; Pettini, Boksenberg, \& Hunstead 1990; Meyer \& York 1992;
Pettini et al. 1994; Wolfe et al. 1994; Smith et al. 1996; although Lu,
Sargent, \& Barlow (1996) challenge this view based on the abundances of
other metals).  By dimming the light of the background sources, dust
reduces the number of QSOs above a given magnitude threshold (Ostriker \&
Heisler 1984; Fall \& Pei 1988, 1993), thus opposing the effect of the
magnification bias due to lensing. It is thus a relevant issue to consider
these two effects simultaneously and to estimate their relative importance.
This is the purpose of the present paper. The detection of lensing by
low-redshift spiral galaxies, such as B0218+357 at $z=0.68$ (O'Dea et al.
1992; Patnaik et al.  1993), 2237+0305 at $z=0.04$ (Huchra et al.
1995), and B1600+434 (Jaunsen \& Hjorth 1997)
implies that dust obscuration does not mask lensing altogether.


The efficiency of gravitational lensing is expected to strongly peak in
the redshift interval $0.2\la z\la 0.8$. Aside from the tendency of
cosmological lenses to maximize their influence around $z\sim 0.5$
(Schneider, Falco, \& Ehlers 1992; Bartelmann \& Loeb 1996), evolutionary
phenomena such as the hierarchical merging of halos and the HI consumption
by star formation, would tend to reduce the significance of lensing per HI
column density at high redshifts.  We therefore focus in this paper only on
spiral lenses at low-redshifts $z\la 0.8$.  At these redshifts, there are
various indications that the potential wells of spirals did not evolve
significantly due to merging:

\begin{itemize}
\item[1.] Substantial merging or accretion would have resulted
in asymmetric distortion of the isophotes of spiral galaxies.  Direct
observations of the level of lopsidedness in spiral galaxies set an upper
limit on the current accretion rate of companions with a mass of order 10\%
the parent galaxy mass, which is 0.25 companions per Gyr (Zaritzky \& Rix
1996). This implies that the fractional increase in the mass of galactic
disks due to major mergers is smaller than 15\% since $z\sim0.5$.

\item[2.] The thinness and coldness of the stellar distribution in
galactic disks imply that these disks did not encounter substantial
merging since gas was converted into stars in them.  T\'oth \& Ostriker
(1992) have obtained an upper limit of 4\% on the fraction of the Galactic
mass inside the solar radius that could have accreted within the last 5
billion years, based on the observed scale and Toomre $Q$--parameter of the
Milky-Way disk.

\item[3.] Considerable evolution in the potential wells of spiral galaxies 
since $z\sim 1$ would have resulted in a considerable scatter in their
Tully-Fisher relation due to variations in their formation histories.
Eisenstein \& Loeb (1996) have used Monte Carlo realizations of halo
formation histories in hierarchical models of structure formation to limit
any substantial evolution in the potential wells of spirals since $z\sim1$,
based on the observed tightness in their Tully-Fisher relation.

\item[4.] The statistics of multiply imaged QSOs implies that galactic lenses
(either ellipticals or spirals) did not encounter significant merging 
below $z\sim1$ (Mao \& Kochanek 1994; Rix et al. 1994).

\item[5.] Measurements of the rotation velocities and luminosities 
of spiral galaxies at intermediate redshifts (Vogt et al 1996) imply that
they are kinematically similar to nearby spirals.

\item[6.] Deep imaging and spectroscopy in the field of 3C 336 
(Steidel et al. 1996) show a lack of strong evolution in the population of
galactic metal--line absorbers up to a redshift $z\sim 1$.
\end{itemize}
For redshifts below unity, the observed evolution in $\Omega_{\rm HI}$ is
also very weak (cf. Fig. 8 in Wolfe et al. 1995). Based on all of this
evidence, we adopt a model in which the gravitational potentials and the
HI+dust content of the local population of spiral galaxies had not evolved
significantly between a redshift of $0.6\pm0.2$ and the present time.

The paper is organized as follows. In \S2 we summarize our
parameterization of dust at high redshift, focusing on its relation to
neutral hydrogen in damped Ly$\alpha$ systems. We then examine in \S3
the simple situation of gravitational lensing and dust obscuration of
a single point source by a single lens. We turn to statistical
considerations in \S4, where we detail our method to calculate the
combined effects of gravitational lensing and dust obscuration 
of a population of background QSOs by an intervening population of
damped Ly$\alpha$ absorbers. Finally, \S5 summarizes our conclusions.

\section{Dust Model}

Direct evidence that damped Ly$\alpha$ systems contain dust is
provided by the reddening of background QSOs 
and by the depletion of Cr relative to Zn.
The amount of dust is usually expressed in terms of the dust-to-gas
ratio $k$ by
\begin{equation}
  k \equiv \frac{\tau_B}{N_{21}}\;,
\label{eq:1}
\end{equation}
where $\tau_B$ is the optical depth due to dust in the rest frame $B$
band and $N_{21}$ is the column density of neutral hydrogen in units
of $10^{21}\;{\rm cm^{-2}}$.

The reddening of background QSOs can be estimated by comparing the
spectral indices of quasars with and without damped Ly$\alpha$
absorbers. Typical dust-to-gas ratios in the
damped Ly$\alpha$ systems can then be inferred from this reddening
estimate if a shape for the extinction curve is assumed.

Another way to infer $k$ is provided by measurements of the gas-phase
abundances of Zn, Cr, and other trace elements. The underlying
assumption is that, apart from overall scalings in the metallicity and
dust-to-gas ratio, the relative composition of the interstellar medium
is the same in the damped Ly$\alpha$ systems as it is in the Milky Way
(i.e., the abundance patterns and grain composition are the same in
both). Measuring the ratio Zn/H and the depletion of Cr relative to
Zn, one can estimate the fraction of gaseous Cr that has condensed
into dust grains. Both methods of measuring $k$ yield a typical
dust-to-gas ratio for damped Ly$\alpha$ systems at $z\approx2$ 
that is roughly 10\% of the Milky Way value 
$k\approx0.8$. 

The dust-to-gas ratio evolves with redshift due to various astrophysical
processes, such as conversion of gas into stars, metal enrichment and
inflows or outflows of gas from galaxies. However, since we focus on low
redshift absorbers, we assume no evolution in their dust-to-gas ratio,
\begin{equation}
  {\bar k}(z)\simeq {\bar k}(0)\; .
\label{eq:2}
\end{equation}

The dust optical depth $\tau(z)$ in an absorber at redshift $z$ is
related to the optical depth in the $B$ band $\tau_B$ by
\begin{equation}
  \tau(z) = \tau_B\;\xi\left(\frac{\lambda_B}{1+z}\right)\;,
\label{eq:3}
\end{equation}
where $\xi(\lambda)$ is the ratio of the extinction at wavelength
$\lambda$ to that in the $B$ band. Observations (Mathis 1990; Savage
\& Mathis 1979) indicate that the extinction curve declines with
wavelength roughly as $\xi(\lambda) \propto \lambda^{-1}$. For 
wavelengths not too far from the $B$ band, a good fit is provided
by 
\begin{equation}
  \xi(\lambda) \simeq 0.4 \frac{1}{\lambda}\;\mu \,{\rm m}\;.
\label{eq:4}
\end{equation}

In terms of the dust-to-gas ratio as defined in equation~(\ref{eq:1}),
the optical depth can be written as 
\begin{equation}
  \tau(z) = k(z)\,N_{21}\,\xi\left(\frac{\lambda_B}{1+z}\right)\;.
\label{eq:5}
\end{equation}

The intrinsic flux $S'$ of a source is changed to $S$ if it passes an
intervening galaxy. Gravitational lensing magnifies the flux by a
factor $\mu$, while obscuration by dust reduces the flux by a factor
$\exp(-\tau)$ (Fall \& Pei 1989). The net effect is thus
\begin{equation}
  S = \mu\,\exp(-\tau)\,S'\;.
\label{eq:6}
\end{equation}

\section{Combination of Lensing and Dust Obscuration}

We now examine the combined effect of dust obscuration and
gravitational lensing in the simple situation of a point source which
is lensed and obscured by a single intervening galaxy. For
definiteness, we adopt a source redshift of $z_{\rm s}=3$ and a lens
redshift of $z_{\rm d}\equiv z_{\rm abs}=0.5$ where necessary.

We describe the profile of the neutral hydrogen disk in the galaxy by
the radius where the face-on column density is $N'$,
\begin{equation}
  R_{\rm H}(N') = f_{\rm H}(N')\,R_*\,l^t\;,
\label{eq:7}
\end{equation}
where $l=L/L_*$ is the galaxy luminosity, $t\approx0.4$, and
$R_*\approx11\,h^{-1}\,{\rm kpc}$ is a characteristic radius of an
$L_*$ galaxy. The data by Broeils \& van Woerden (1994) imply an
exponential neutral hydrogen profile,
\begin{equation}
  f_{\rm H}(N') = \frac{3}{2} -
  2\log\left(\frac{N'}{2\times10^{20}\;{\rm cm^{-2}}}\right)\;.
\label{eq:8}
\end{equation}

For simplicity, we will in this section consider a lensing galaxy with
luminosity $L_*$. The hydrogen radius of equation~(\ref{eq:7}) then
simplifies to $R_{\rm H}(N')=f_{\rm H}(N')\,R_*$, where $N'$ is the
face-on column density of HI. If the disk is inclined by an angle
$\theta$ with respect to the line-of-sight, the observed column
density is $N=N'/\cos\theta$.

The intercepted HI column density not only depends on the relative
position of absorber and source, but also on the lensing properties of
the absorber because the light rays are bent. We model the lens as a
singular isothermal sphere (SIS). Recent work (Kochanek 1996) has
shown that, in the case of elliptical galaxy lenses, the addition of a
small core radius, consistent with the kinematics and the image
configuration of the lens, does not change the lensing probability
significantly. Since the light distribution in the central bulge of
spiral galaxies resembles that in elliptical galaxies, we assume the
SIS model to be a good approximation for our purposes.

In the SIS model, the scaled distances of the source $y$ and of the
images $x$ from the center of the lens are related by (e.g.,
Schneider et al. 1992)
\begin{equation}
  x = \left\{
  \begin{array}{ll}
    y\pm1 & \hbox{if $y\le1$} \\
    y+1   & \hbox{if $y>1$}   \\
  \end{array}\right.\;.
\label{eq:9}
\end{equation}
The appropriate length scale for $x$ is
\begin{equation}
  \zeta_0 \equiv 4\pi\left(\frac{\sigma_v}{c}\right)^2\,
  \frac{c}{H_0}\,\frac{r_{\rm d}\,r_{\rm ds}}{r_{\rm s}}\;,
\label{eq:10}
\end{equation}
where $r_{\rm d,s,ds}$ are the angular diameter distances from the
observer to the lens, to the source, and from the lens to the source,
respectively, in units of the Hubble length $c\,H_0^{-1}$. In the
numerical examples throughout the paper, we will use a Hubble constant
$H_0=100\,h\;{\rm km\,s^{-1}\,Mpc^{-1}}$, with $h=0.5$, and an $L_*$
galaxy velocity dispersion of
$\sigma_{v,*} \approx 160\;{\rm km\,s^{-1}}$.
The magnification of 
a point source by 
an SIS lens is
\begin{equation}
  \mu = \left\{
  \begin{array}{ll}
    {2/y} & \hbox{if $y\le1$} \\
    1 + {1/y} & \hbox{if $y\ge1$} \\
  \end{array}\right.\;.
\label{eq:16}
\end{equation}

In order to compute the optical depth due to dust obscuration, we need
to calculate the column density $N'$ of the galactic HI disk where it
is intercepted by the light rays from a source at position $y$. Since
an inclined disk appears elliptical, the radius in the plane of the
disk where the light rays pass the galaxy depends on the relative
position angle between $y$ and the inclined galactic disk. For an
inclination angle $\theta$, the radius averaged over position angle is
\begin{equation}
  \langle x\rangle = \frac{1}{2\pi}\,\int_0^{2\pi}\,d\varphi\,
  \left[(x\cos\varphi)^2 + (x\,\sin\varphi/\cos\theta)^2\right] =
  \frac{2x}{\pi\cos\theta}\,{\rm E}(\sin^2\theta)
\label{eq:12}
\end{equation}
with the elliptical function of the $2^{\rm nd}$ kind ${\rm E}(x)$.

The HI column density $N'$ in the disk at the average radius $\langle
x\rangle$ is then found from
\begin{equation}
  R_*\,f_{\rm H}(N') = \langle x\rangle\,\zeta_0
\label{eq:13}
\end{equation}
which can be solved for $N'$ using $f_{\rm H}(N')$ from
equation~(\ref{eq:8}). The observed column density $N$ as a function
of the source position $y$ is
\begin{equation}
  N(y) = \frac{2\times10^{20}\;{\rm cm^{-2}}}{\cos\theta}
  10^{3/4-\langle x(y)\rangle\zeta_0/(2R_*)}\;,
\label{eq:14}
\end{equation}
with $x(y)$ given by equation~(\ref{eq:9}). The optical depth can then be
easily computed from equation~(\ref{eq:5}).  For the redshifts chosen,
$z_{\rm s}=3$ and $z_{\rm d}=0.5$, for observations in the B-band, and for
a cosmological model with $\Omega_0=0.2$ , $\Omega_{\lambda}=0$, we find from
equations~(\ref{eq:4}) and (\ref{eq:10})
\begin{equation}
  \zeta_0 \approx 4\;{\rm kpc}\;,\quad
  \xi \approx 1.4\;,\quad
  R_* = 22\;{\rm kpc}\;,
\label{eq:15}
\end{equation}
for 
$h=0.5$.

\begin{figure}[t]
\centerline{\epsfxsize=0.6\hsize\epsffile{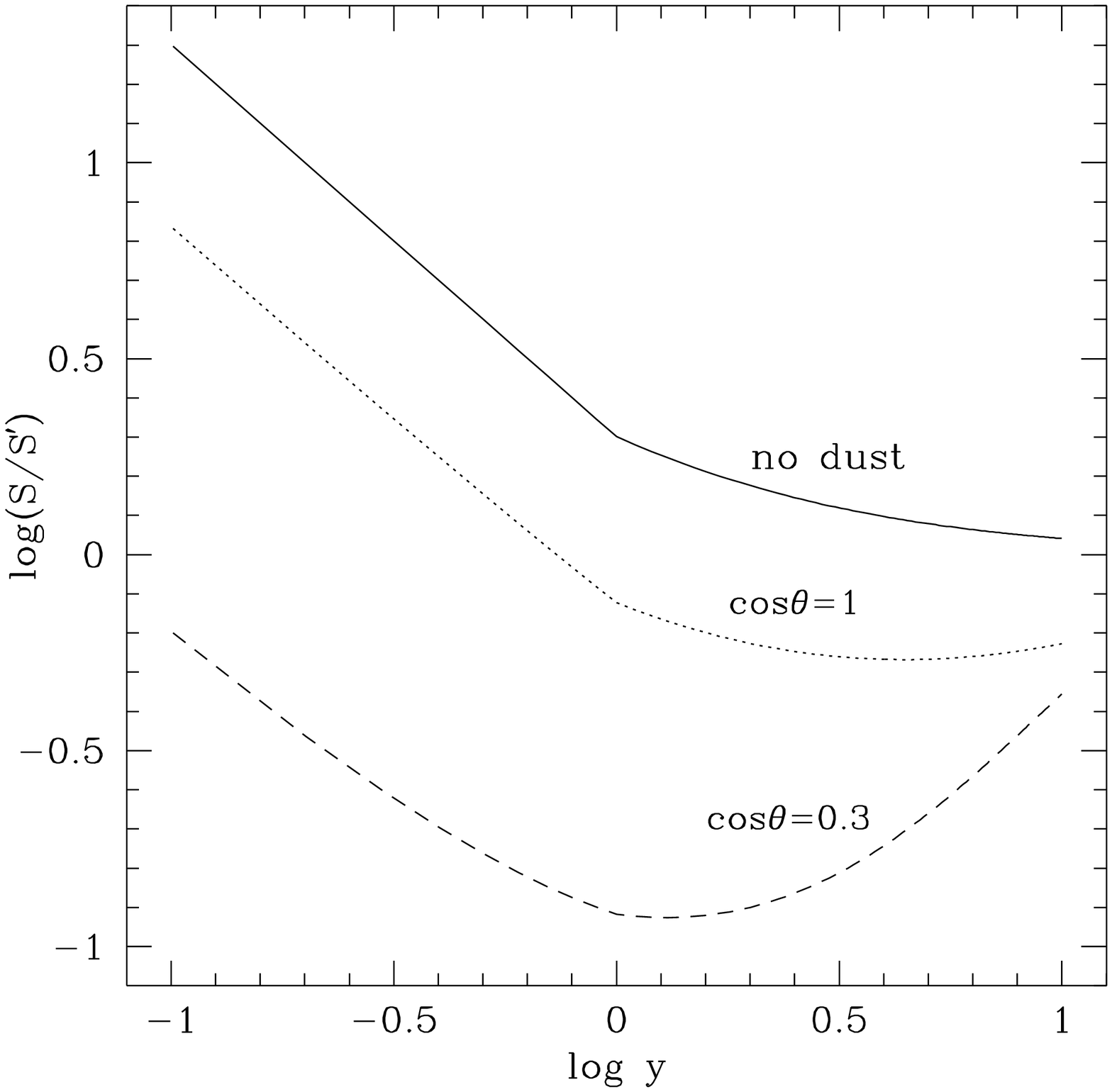}}
\caption{The ratio between the observed flux $S$ and the intrinsic QSO
flux (without the 
absorbing lens), 
$S^\prime$, as a function of the dimensionless impact parameter $y$.
The solid curve shows the ratio where only lensing is taken into
account, while the other two curves consider the combined effect of
lensing and dust for a face-on disk (dotted line) and a disk inclined
by $70^\circ$ (dashed line).} 
\label{fig:1}
\end{figure}

We now have all the ingredients to estimate the ratio between the
observed and intrinsic fluxes, $S$ and $S'$, related by
equation~(\ref{eq:6}). The resulting flux ratio $S/S'$ is shown in
Figure~\ref{fig:1}, where $S/S'$ is plotted as a function of source
position $y$ for different values of the inclination angle $\theta$ of
the disk. For comparison, the solid line shows $S/S'$ without taking
dust obscuration into account.
Figure \ref{fig:1} shows that for each inclination angle there is a
source position $y_*$ such that magnification dominates over dimming
(i.e., $S/S'>1$) for $y<y_*$ while extinction is stronger than
magnification (i.e., $S/S'<1$) for $y>y_*$. As smaller $y$ imply
smaller $x$, we find that magnification is stronger than extinction
for light rays passing closer to the center of the disk.

\section{Statistics of Dust Obscuration and Lensing}

We now examine the combined effect of lensing and dust obscuration on
a population of distant QSOs by a population of intervening galaxies.
Damped Ly$\alpha$ systems offer one of the best ways of tracing the
evolution of neutral hydrogen in spiral galaxies at intermediate and
high redshifts. We therefore focus on the simultaneous effects of 
lensing and dust obscuration on the properties of neutral hydrogen
which can be inferred from the damped absorption lines. In particular,
we investigate the effects of dust and lensing on the column density
distribution function of HI, $f(N)$, and on the estimates of the
contribution of HI to the density of the Universe $\Omega_{\rm HI}$.

We model the column density profile of neutral hydrogen in the
absorbers according to equations~(\ref{eq:7}) and (\ref{eq:8}) and
assume for simplicity that it is constant in time. The absorber
population is described by a Schechter luminosity function,
\begin{equation}
  \Phi(l)\,dl = \Phi_*\,l^s\,\exp(-l)\,dl\;,
\label{eq:17}
\end{equation}
for which we adopt the parameters
appropriate for spiral galaxies (Marzke et al. 1994),
\begin{equation}
  \Phi_* = 1.5\times10^{-1}\,h^2\;{\rm Mpc^{-3}}\;,\quad
  s = -0.81.
\label{eq:18}
\end{equation}

The source population of distant QSOs is parameterized in terms of the
intrinsic number count function which can be well modeled by a broken
power law (e.g., Boyle, Shanks, \& Peterson 1988; Hartwick \& Schade
1990),
\begin{equation}
  \frac{dN}{dS}(S) = A\times \left\{
  \begin{array}{ll}
    S^{-\beta_1-1} & \hbox{if $S\le S_0$} \\
    S^{-\beta_2-1} & \hbox{if $S>S_0$}    \\
  \end{array}\right.\;,
\label{eq:19}
\end{equation}
where the flux $S_0$ at the break point corresponds to an apparent QSO
magnitude of $B\approx19.5$ and $A$ is a constant.  A reasonable fit
to observational data for the two branches of the power law is
provided by the values $\beta_1=0.64$ and $\beta_2=2.52$ (Pei 1995).

If neither lensing nor dust obscuration are effective, the probability
for a QSO to exhibit damped Ly$\alpha$ absorption with a column
density $\ge N$ in its spectrum is given by (Bahcall \& Peebles 1969)
\begin{equation}
  P_{{\rm Ly}\alpha}(z_1,z_2,N) = \Phi_*\,\Gamma(1+s+2t)\pi\,
  R^2_*\,\langle f^2_{\rm H}(N)\rangle\,\Delta X(z_1,z_2)\;,
\label{eq:20}
\end{equation}
where $\Delta X(z_1,z_2)$ is the absorption distance surveyed per QSO,
and $\langle f^2_{\rm H}(N)\rangle$ is the average of (\ref{eq:8})
over all inclination angles.

If the absorption probability $P_{{\rm Ly}\alpha}(z_1,z_2,N)$ is
measured, the column density distribution of absorbers $f(N)$ is given
by
\begin{equation}
  f(N) = \frac{c}{H_0}\,\frac{1}{\Delta X(z_1,z_2)}\,
  \left|\frac{\partial P(z_1,z_2,N)}{\partial N}\right|\;.
\label{eq:21}
\end{equation}
This expression is independent of the QSO redshift, the QSO flux, and
the redshift interval surveyed for damped Ly$\alpha$ absorption.

On the other hand, if gravitational lensing is taken into account, the
probability for damped Ly$\alpha$ absorption, and thus the column
density distribution, does depend on the flux threshold $S$ and on the
redshifts involved because of the magnification bias (Bartelmann \&
Loeb 1996). Specifically,
\begin{equation}
  f_{\rm GL+dust}(z_1,z_2,z_{\rm s},N;S) =
  \frac{c}{H_0}\,\frac{1}{\Delta X(z_1,z_2)}\,
  \int_{z_1}^{z_2}\,dz\,{\tilde f}_{\rm GL+dust}(z,z_s,N;S)
\label{eq:29}
\end{equation}
with
\begin{eqnarray}
  {\tilde f}_{\rm GL+dust}(z,z_{\rm s},N;S) &=&
  \frac{1}{N_{\rm QSO}(S)} 
  \int_0^\infty\,dS'\left|\frac{
  \partial p'_{\rm GL}\left(z,z_{\rm s},N;S/S'\right)}
  {\partial N}\right|
  \frac{dN_{\rm QSO}}{dS}(S') \nonumber\\
\label{eq:22}
\end{eqnarray}
where $p'_{\rm GL}(z,z_{\rm s},N;\mu)$ is the probability that a
QSO at redshift $z_{\rm s}$ shows damped Ly$\alpha$ absorption with
column density $\ge N$ and is magnified by a factor $\ge\mu$ by lenses
in the redshift interval $dz$ around $z$ (cf.~equation~(2.25) in
Bartelmann \& Loeb 1996).

Let us now consider the combined effect of lensing and dust obscuration. 
For an intrinsic source flux $S'$, the observed
flux is $S=S'~\mu~\exp(-\tau)$ with
$\tau(k,z)=kN_{21}\xi({\lambda_B/ {1+z}})$ (cf.
eqs.~(\ref{eq:5}) and (\ref{eq:6})).
Consequently, the distribution
of absorbers in equation~(\ref{eq:22}) becomes
\begin{eqnarray}
 {\tilde f}_{\rm GL+dust}(z,z_{\rm s},S;N) &=& 
\frac{1}{N_{\rm QSO}(S)}\,
 \int_0^\infty\,dS'\left|\frac{
\partial p'_{\rm GL}\left(z,z_{\rm s},N;S/S'e^{-\tau}\right)}
  {\partial N}\right|\frac{dN_{\rm QSO}}{dS'}(S') \nonumber\\
  &=& \frac{A}{N_{\rm QSO}(S)}\,\left[\int_0^{S_0}\,dS'\left|\frac{
   \partial p'_{\rm GL}\left(z,z_{\rm s},N;S/S'e^{-\tau}\right)}
   {\partial N}\right|S'^{-\beta_1-1}\right. \nonumber\\
  &+& \left. \int_{S_0}^\infty\,dS'\left|\frac{
    \partial p'_{\rm GL}\left(z,z_{\rm s},N;S/S'e^{-\tau}\right)}
    {\partial N}\right|S'^{-\beta_2-1}\right], \nonumber\\
\label{eq:40}
\end{eqnarray}
where we have used equation~(\ref{eq:19}).
By changing 
variables ${\tilde S}=S'e^{-\tau}$, equation~(\ref{eq:40})
becomes
\begin{eqnarray}
 {\tilde f}_{\rm GL+dust}(z,z_{\rm s},S;N) &=& \frac{A}{N_{\rm
QSO}(S)}\,\left[e^{-\beta_1\tau}\int_0^{S_0e^{-\tau}}\,d{\tilde S}
\left|\frac{ \partial p'_{\rm GL}\left(z,z_{\rm s},N;S/{\tilde
S}\right)} {\partial N}\right|{\tilde S}^{-\beta_1-1}\right. \nonumber\\
&+& \left. e^{-\beta_2\tau}\int_{S_0e^{-\tau}}^\infty\,d{\tilde
S}\left|\frac{ \partial p'_{\rm GL}\left(z,z_{\rm s},N;S/{\tilde
S}\right)} {\partial N}\right|{\tilde S}^{-\beta_2-1}\right].\nonumber\\
\label{eq:41}
\end{eqnarray}    
Hence, dust causes an apparent shift in the break flux of the QSO
distribution function and damps the absorber distribution by a factor ${\rm
exp}(-\beta\tau)$ (a discussion of the latter effect for bright QSOs can be
found in Fall \& Pei 1993).
 
\begin{figure}[t]
\centerline{\epsfysize=3.7in\epsffile{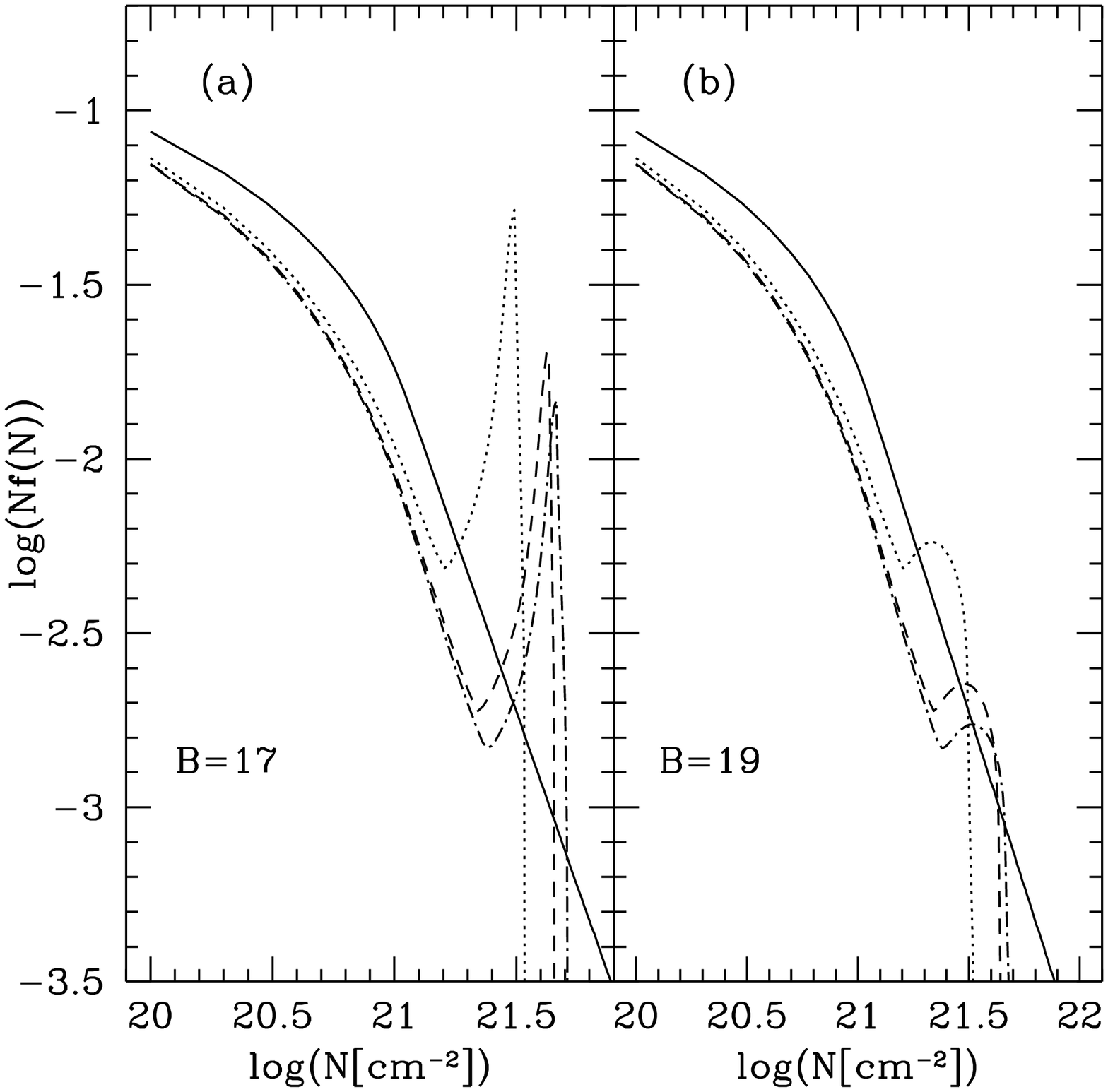}}
\caption{Column density distribution of damped Ly$\alpha$ absorbers
per logarithmic column density interval, $Nf(N)$, for samples of QSOs
with flux limit $B=17$ [panel(a)] and $B=19$ [panel(b)].   
The solid curve in
both panels displays $Nf(N)$ without the influence of 
gravitational lensing and dust extinction. 
The other three curves take into account both lensing and dust 
for three cosmological models characterized respectively by the parameters
$\Omega_0$=0.2, $\Omega_{\lambda}$=0.8 (dotted line); $\Omega_0$=0.2,
$\Omega_{\lambda}$=0 (dashed line); and $\Omega_0$=1,
$\Omega_{\lambda}$=0 (dashed-dotted line).w 
In both panels the
QSO redshift is fixed at $z_{\rm s}=3$ and the redshift interval
for the absorbers is $0.4\leq z_{\rm abs}\leq 0.8$}
\label{fig:2}
\end{figure}

Since the value of $k$ varies considerably among different DLA systems, we
allow for a spread in its values. Following Fall \& Pei (1993), we assume 
a log-normal form for the probability distribution of $k$,
\begin{equation}
  g(k) = \frac{1}{\sqrt{2\pi}\sigma_*\,k}
  \exp\left[-\frac{(\ln k-\ln k_*)^2}{2\sigma_*^2}\right]\;,
\label{eq:26}
\end{equation}
where the parameter $k_*$ is related to the mean
$\bar k$  of the dust-to-gas ratio by
\begin{equation}
  \bar k \equiv \int_0^\infty\,dk\,g(k)\,k =
  k_*\,\exp\left(\frac{\sigma_*^2}{2}\right)\,,
\label{eq:27}
\end{equation}
with $\bar k$ given by equation~(\ref{eq:2}) under the assmption that
at $z=0$ the Milky Way value of $k$ reflects the true mean
of the spiral galaxy population. Following the study of
Fall \& Pei (1993; cf. their figure 8b),we adopt a
constant value of $\sigma_*=1$.   With the
introduction of a probability distribution for $k$, equation~(\ref{eq:41})
reads
\begin{eqnarray}
 {\tilde f}_{\rm GL+dust}(z,z_{\rm s},N,S) &=& 
  \frac{A}{N_{\rm QSO}(S)}\;\int_0^\infty dkg(k,z) \nonumber\\
  &\times&\left[
    e^{-\beta_1\tau(k,z)}\,\int_0^{S_0e^{-\tau(k,z)}}\,dS'\,
    \frac{\partial p'_{\rm GL}\left(z,z_{\rm s},N,S/S'\right)}
    {\partial N}
    S'^{-\beta_1-1}\right.\nonumber\\
  &+&\left.
    e^{-\beta_2\tau(k,z)}\,\int_{S_0e^{-\tau(k,z)}}^\infty\,dS'\,
    \frac{\partial p'_{\rm GL}\left(z,z_{\rm s},N,S/S'\right)}
    {\partial N}
    S'^{-\beta_2-1}\right]\;.\nonumber\\
\label{eq:30}
\end{eqnarray}
Equations~(\ref{eq:29}) and ~(\ref{eq:30}) can be used to get the observed
distribution of absorbers in the redshift interval $z_1\leq z_{\rm abs}\leq
z_2$. The results from these equations are displayed in Figure~\ref{fig:2}
for the absorber redshift interval $0.4\leq z_{\rm abs}\leq 0.8$. This
figure shows the observed distribution of $N\,f(N)$ for two choices
of the limiting flux $B$ and for three cosmological models, in comparison
to the true distribution (solid line).

\begin{figure}[t]
\centerline{\epsfysize=3.5in\epsffile{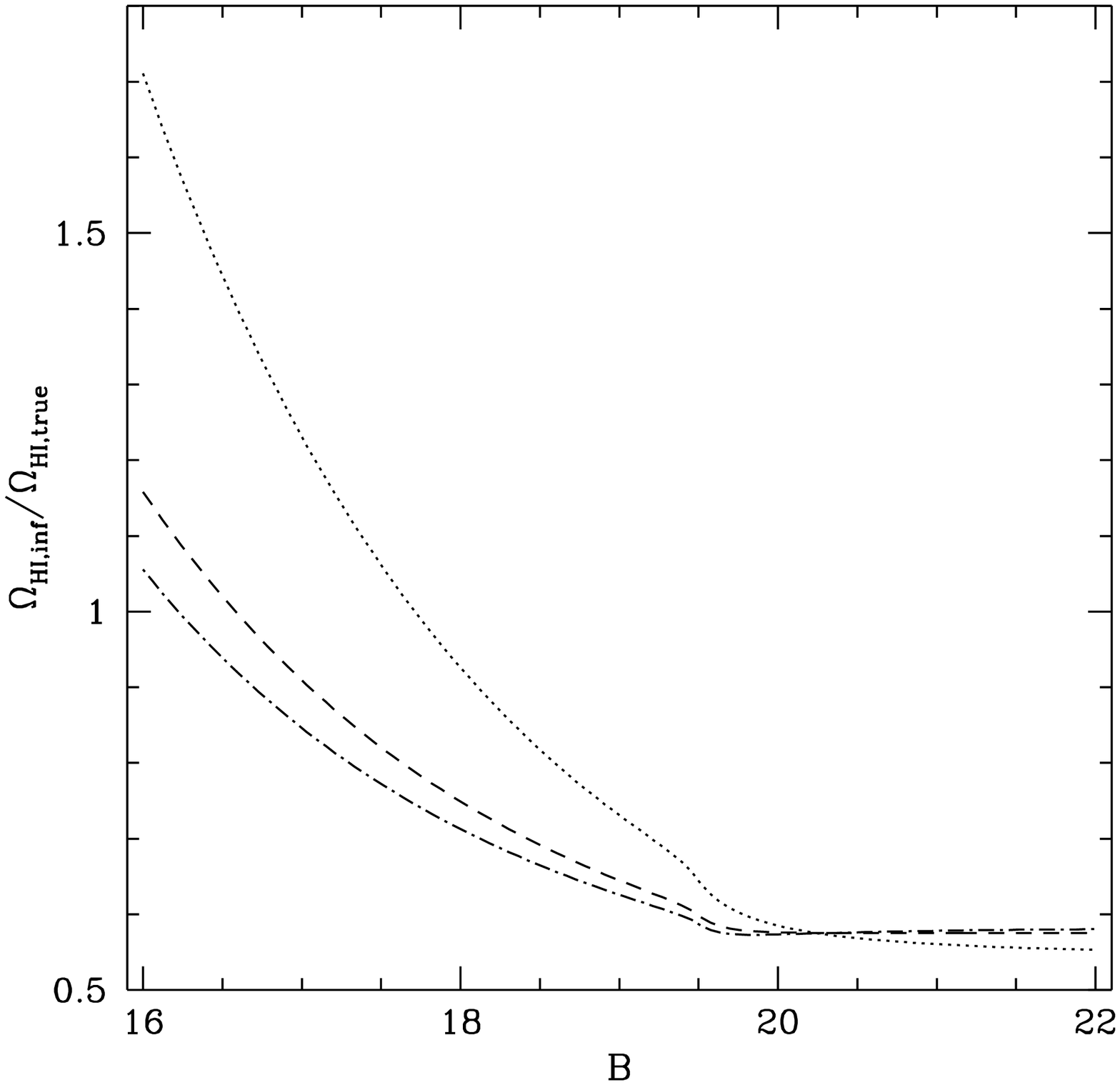}}
\caption{The ratio between the inferred and the true $\Omega_{\rm HI}$
for three cosmological models characterized respectively by the parameters
$\Omega_0$=0.2, $\Omega_{\lambda}$=0.8 (dotted line); $\Omega_0$=0.2,
$\Omega_{\lambda}$=0 (dashed line); and $\Omega_0$=1,
$\Omega_{\lambda}$=0 (dashed-dotted line).} 
\label{fig:3}
\end{figure}

Having calculated $Nf(N)$, the inferred column density of neutral
hydrogen in absorbers with column density of $N_1\le N\le N_2$,
normalized by the present critical density, is given by
\begin{equation}
  \Omega_{\rm HI}(N_1,N_2) = 
  \frac{H_0}{c}\,\frac{\bar m}{\rho_{\rm c,0}}
  \int_{N_1}^{N_2}\,dN\,Nf(N)\;,
\label{eq:31}
\end{equation}
where $\bar m$ is the mean molecular mass per proton and $\rho_{\rm
c,0}$ is the present-day closure density,
\begin{equation}
  \rho_{\rm c,0} = \frac{3H_0^2}{8\pi G}\;.
\label{eq:32}
\end{equation}
Figure~\ref{fig:3} displays the result for $\Omega_{\rm HI}$ for the
absorber redshift interval $0.4\le z_{\rm abs}\le 0.8$ and for the same
three cosmological models as in Figure~\ref{fig:2}.

The enhancement in the abundance of absorbers (and hence in the inferred
density parameter of neutral hydrogen) is a consequence of the
magnification bias due to lensing, which increases the number of QSOs above
the detection threshold.  Extinction by dust results in an opposite
tendency.  In Figure 4a we show the combined effect of these opposite
tendencies as a function of the QSO magnitude limit, for low-redshift
absorbers with $N\ga 10^{20}\rm{cm^{-2}}$. The QSO detection probability is
compared to the case where neither of these two effects are taken into
account.  We find that for bright quasars with a limiting magnitude $B\la
17$, the effect of lensing dominates.  In Figure 4b we compare the above
probability to the situation where only lensing is taken into account (cf.
Bartelmann \& Loeb 1996). This figure illustrates that for low-redshift
absorbers, the suppression of the lensing effect by dust is of several tens
of percent.  The level of suppression is higher for brighter QSOs because
of the dependence of the extinction factor on the slope of the QSO number
counts [cf.  eq.~(\ref{eq:30})], which is steeper for brighter QSOs.  Note
that the suppression of the lensing probability by dust is weaker than the
suppression of the lensing effect on $\Omega_{\rm HI}$ (compare Fig. 5 of
Bartelmann \& Loeb (1996) to Fig.~\ref{fig:3} in this paper). This follows
from the fact that $\Omega_{\rm HI}$ is integrated with an extra factor of
$N$ (cf. eq.~\ref{eq:31}), which tends to emphasize high column density
systems more, and therefore enhance the suppression of lensing by dust at a
given QSO magnitude limit.

\section{Conclusions}

Gravitational lensing by a spiral galaxy occurs when the line-of-sight
to a background quasar passes within a few kpc from the center of the
galactic disk. Since galactic disks are rich in neutral hydrogen, the
quasar spectrum will likely be marked by a damped \lya absorption
trough at the lens redshift. Indeed, the observed 21 cm absorption in
the spiral galaxy lens B0218+357 implies an HI column density of
$4\times10^{21}~{\rm cm^{-2}}\,(T_{\rm s}/100~{\rm K})/(f/0.1)$, where
$T_{\rm s}$ is the spin temperature and $f$ is the HI covering factor
in this system\footnote{The high HI column density in
this lens will be directly calibrated based on the absorption
spectrum of the QSO 
in forthcoming HST observations (Falco, Loeb, \&
Bartelmann 1996).}.
Bartelmann \& Loeb (1996) have shown that the
efficiency of searches for gravitational lensing with sub-arcsecond
splitting can be enhanced by 1-2 orders of magnitude if one focuses on
the subset of bright quasars which show low-redshift ($z_{\rm abs}\la
0.8$) strong \lya absorption ($N\ga 10^{21}~ {\rm cm^{-2}}$). The
double-image signature of lensing could, in principle, be identified
spectroscopically and without the need for high-resolution imaging;
the absorption spectrum of the lens might show a generic double-step
profile due to the superposition of the two absorption troughs of the
different images.  The different images pass the absorbing disk at
different impact parameters and therefore have different spectral
widths.

Lensing of optical sources by spiral galaxies is expected to be
counteracted by dust obscuration, since the central opacity of nearby
disks is considerable (see, e.g. Byun 1993; Davies et al. 1993; Byun,
Freeman, \& Kylafis 1994; Jansen et al. 1994; Rix 1995). The
distribution of dust in galactic disks is correlated with the
distribution of neutral hydrogen, which in turn can be probed in the
absorption spectra of QSOs.  In this paper, we used the estimated
dust-to-gas ratio of our galaxy and allowed a spread in its values
to calculate the effects of dust on lensing by low redshift spiral
galaxies. 
For absorbers with $z_{\rm abs}\la0.8$, we have found that the
relative importance of the magnification bias due to lensing and of the
obscuration due to dust depends on the QSO magnitude limit and 
 on the choice of cosmological parameters (cf. Fig.~4a).
Overall, the effect of dust suppresses the detection probability of lensing
by $\la 55\%$ (cf.~Fig.~4b).

The magnification bias and the deflection of light rays due to lensing
change the column density distribution of damped \lya absorbers, $f(N)$.
Although dust tends to suppress the distortion to $f(N)$ derived by
Bartelmann \& Loeb (1996), the generic lensing peak for low-redshift
absorbers with $N \sim
10^{21.5} \rm{cm^{-2}}$ is still evident in Figure~2a. This 
peak results from the magnification bias, which raises faint QSOs above the
detection threshold, thereby enhancing the probability for observing damped
\lya absorbers.  The sharp cutoff at yet higher column densities arises
because the bending of light rays due to lensing yields a minimum
value for the impact parameter of the line-of-sight to the QSO through
the galactic disk. For low-redshift absorbers and models with no
cosmological constant we have found that the density parameter of
neutral hydrogen $\Omega_{\rm HI}$, obtained by integrating $Nf(N)$
over all column densities, is overestimated for QSO samples with $B\la
16$ and is underestimated for fainter QSOs (cf. Fig.~3).

Since the current set of known damped \lya systems includes only a small
number of low-redshift absorbers (see, e.g. Table 3 in Wolfe et al.
(1995)), it is difficult to compare the statistical results shown in
figures 2--4 to existing data.  We expect the significance of lensing
relative to dust to decline at $z\ga1$, and to yield only a secondary
correction to the inferred statistics of the current set of damped \lya
absorbers (Wolfe et al.  1995; Storrie-Lombardi et al.  1996a,b). 
The quantitative predictions of this paper will be tested in a
straightforward manner with the advent of absorption data for larger
statistical samples of QSOs with uniform selection criteria, such as the
set of $\sim 10^5$ QSOs in the forthcoming Sloan digital sky survey (Gunn
\& Knapp 1993).

\acknowledgements

The authors wish to thank an anonymous referee for insightful
comments.  This work was supported in part by the NASA ATP grant
NAG5-3085 (for AL), and by the Sonderforschungsbereich SFB 375-95 of
the Deutsche Forschungsgemeinschaft (for MB).

\begin{figure}[t]
\centerline{\epsfysize=3.7in\epsffile{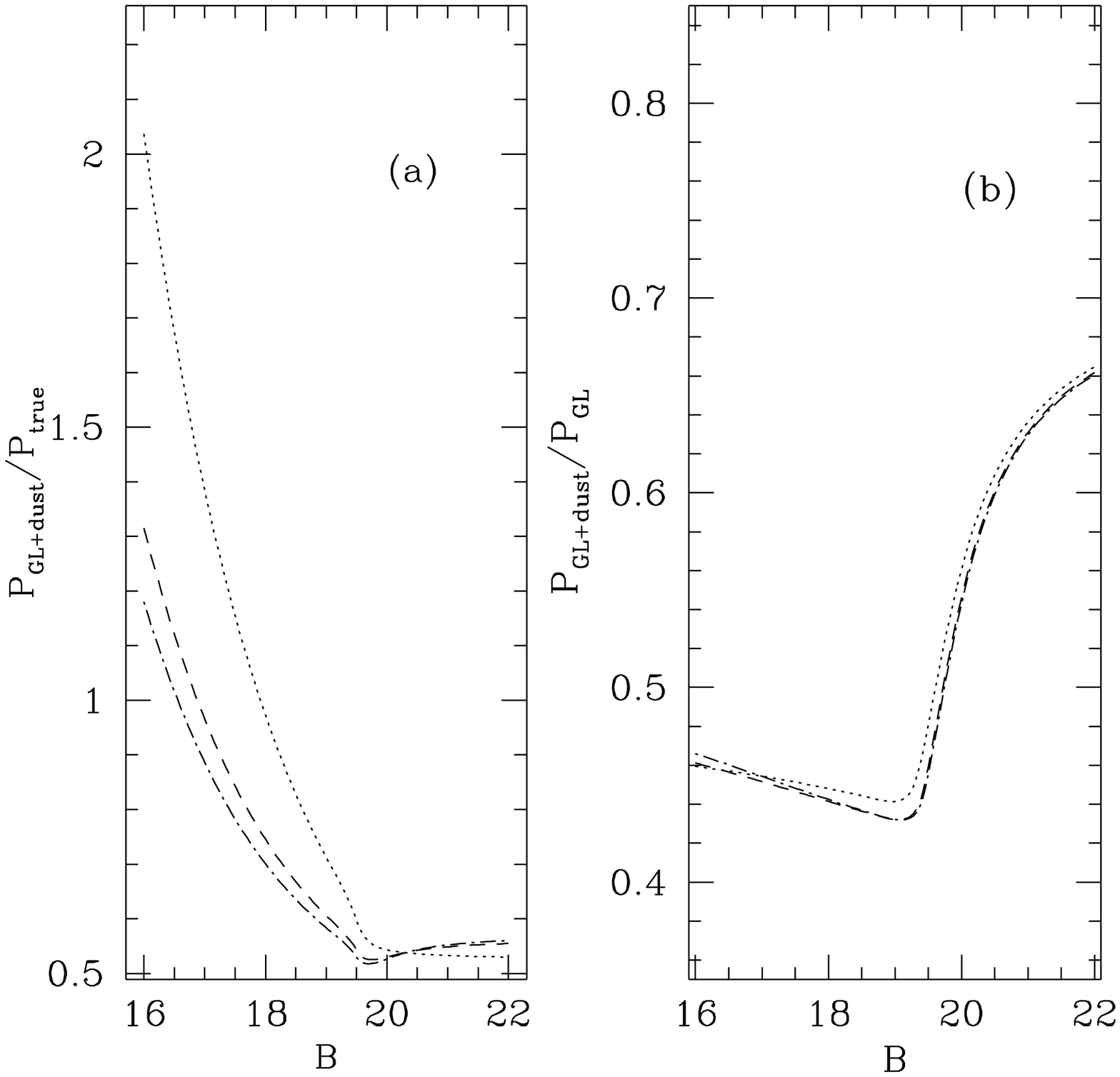}}
\caption{Probability for a QSO to be detected with 
apparent 
magnitude brighter than $B$ and show damped Ly$\alpha$ absorption
with 
column density $N>10^{20}\;{\rm cm^{-2}}$ when lensing and dust
obscuration are taken into account. The probability is normalized by
its value 
in the absence 
of an absorber [panel (a)], or with only lensing taken into account
[panel (b)]. In both panels, $z_{\rm s}=3$, $0.4\leq z_{\rm
abs}\leq0.8$, and the three curves refer to the same models as in
Figures \protect\ref{fig:2} and \protect\ref{fig:3}.}
\label{fig:4}
\end{figure}

\end{document}